\documentclass[reprint,aps,prb,superscriptaddress,showpacs,footinbib,citeautoscript,noeprint,twocolumn]{revtex4-1}
\usepackage{float}
\usepackage{graphicx}
\usepackage{color}
\usepackage{amsmath}
\usepackage{amssymb}
\usepackage{subcaption}
\usepackage{ragged2e}
\usepackage{braket}
\usepackage{siunitx}
\usepackage{mathtools}
\DeclareCaptionJustification{justified}{\justifying}
\usepackage[justification=justified, format=plain]{caption}
\usepackage{booktabs}
\DeclareGraphicsExtensions{.eps,.png,.jpg,.pdf}

\begin{document}
\title[]{\textit{Ab initio} calculation of the spin lattice relaxation time $T_1$ for nitrogen-vacancy centers in diamond}
\author{J.Gugler}
\affiliation{Institute of Applied Physics, TU Wien, Wiedner Hauptstr. 8-10/134, 1040 Vienna, Austria}
\email{jg@cms.tuwien.ac.at}
\author{T.Astner}
\affiliation{Vienna Center for Quantum Science and Technology, Atominstitut, TU Wien, Stadionallee 2, 1020 Vienna, Austria}
\author{A. Angerer}
\affiliation{Vienna Center for Quantum Science and Technology, Atominstitut, TU Wien, Stadionallee 2, 1020 Vienna, Austria}
\author{J. Schmiedmayer}
\affiliation{Vienna Center for Quantum Science and Technology, Atominstitut, TU Wien, Stadionallee 2, 1020 Vienna, Austria}
\author{J.Majer}
\affiliation{Vienna Center for Quantum Science and Technology, Atominstitut, TU Wien, Stadionallee 2, 1020 Vienna, Austria}
\author{P. Mohn}
\affiliation{Institute of Applied Physics, TU Wien, Wiedner Hauptstr. 8-10/134, 1040 Vienna, Austria}
\begin{abstract}
We investigate the fundamental mechanism of spin phonon coupling in the negatively charged nitrogen vacancy center ($\mathrm{NV}^-$) in diamond in order to calculate the spin lattice relaxation time $T_1$ and its temperature dependence from first principles. Starting from the dipolar spin-spin interaction between two electrons, we couple the spins of the electrons to the movements of the ions and end up with an effective spin-phonon interaction potential $V_{\mathrm{s-ph}}$. Taking this time dependent potential as a perturbation of the system, a Fermi's golden rule expression for transition rates is obtained which allows to calculate the spin lattice relaxation time $T_1$. \textcolor{black}{We find that the temperature dependence of $T_1$ is determined by the the zero temperature transition rate $\Gamma_0$.}  We simulate the color center \textit{ab initio} to extract the figures necessary to quantify $\Gamma_0$. We calculate the local phonon modes of the color center within the harmonic approximation using the small displacement method and extract the phononic density of states and bandstructure by diagonalizing the dynamical matrix. We show that our model allows to calculate $T_1$ in good agreement with experimental observations.
\end{abstract}
\maketitle
\section{Introduction}
\label{sec:Intro}
The negatively charged nitrogen vacancy center ($\mathrm{NV}^-$) is an important colour center in diamond \cite{Doherty2013} that consists of a substitutional nitrogen atom adjacent to a vacant lattice site. Six electrons are located at the center, which exhibits $C_{\mathrm{3v}}$-symmetry, and they form an electronic ground state spin triplet transforming according to the $A_2$ representation. This state is further split by the dipolar spin-spin interaction into a $m_\mathrm{s} = 0$ ground state and two degenerate excited $m_\mathrm{s} = \pm 1$ states with a zero field splitting constant $D/h=\SI{2.88}{\giga\hertz}$ \cite{Acosta2010,Ivady2014}. The spin of the system can be prepared and read out optically \cite{Gruber1997} which leads to many applications in magnetometry \cite{Balasubramanian2008,Rondin2013,Hong2013,Barry2016}, biolabelling \cite{Fu2007}, nano-sensing \cite{Dolde2011,Neumann2013} and makes it a promising candidate for a solid state quantum bit \cite{Wrachtrup2001,Nizovtsev2005,Robledo2011}. Since the spin is the quantity to be manipulated in applications, a proper understanding of spin relaxation is of utmost importance. In this paper we deal with the longitudinal spin relaxation in the $^3A_2$ ground state triplet caused by the interaction of the electron spins with the phonons of the crystal.\\
Experimental studies \cite{Redman1991,Jarmola2012} have suggested that the temperature dependence of the spin-lattice relaxation rate in a range between \SIrange{10}{500}{\kelvin} is well described by a two-phonon Raman process and an Orbach process \cite{Orbach1961}, however there are measurements where a different behaviour was observed \cite{Takahashi2008}. Also, the measured relaxation rates differ by one order of magnitude for different samples. To understand and predict spin-lattice relaxation times quantitatively in this system the fundamental mechanism of spin-phonon coupling has to be investigated. Insight into this coupling mechanism is most easily achieved by considering first-order processes, which are dominant at low temperatures around the spin transition energy $D$ of the spins ($T = \SI{138}{\milli\kelvin}$). At these temperatures the phonon spectrum is frozen out and thus higher-order processes are suppressed. In a recent paper \cite{Astner2017}  a direct single phonon relaxation process and spin lattice relaxation times $T_1$ of up to \SI{8}{\hour} in this temperature regime were observed using a cavity QED protocol.
Since no higher order processes were observed, the measured data are suitable to obtain a fundamental understanding of the spin-phonon coupling mechanism in this system.
\\
This paper is organized as follows: In Sec.~\ref{sec:Theory} we derive an effective spin-phonon interaction $V_{\mathrm{s-ph}}$ starting from the dipolar spin-spin interaction between two electrons and we give an expression for the spin-lattice relaxation rate $\Gamma_1$. In Sec.~\ref{sec:Methods} we explain the computational methods used to calculate $\Gamma_1$ \textit{ab initio} by modelling both the electronic and phononic properties of the center by means of density functional theory. In Sec.~\ref{sec:Results} the influence of lattice defects on $\Gamma_1$ is investigated and a comparison of our results with experimental data is presented followed by the conclusion in Sec.~\ref{sec:Conclusion}.
\section{Theory}
\label{sec:Theory}
\begin{figure}
	 \begin{center}
	 \includegraphics{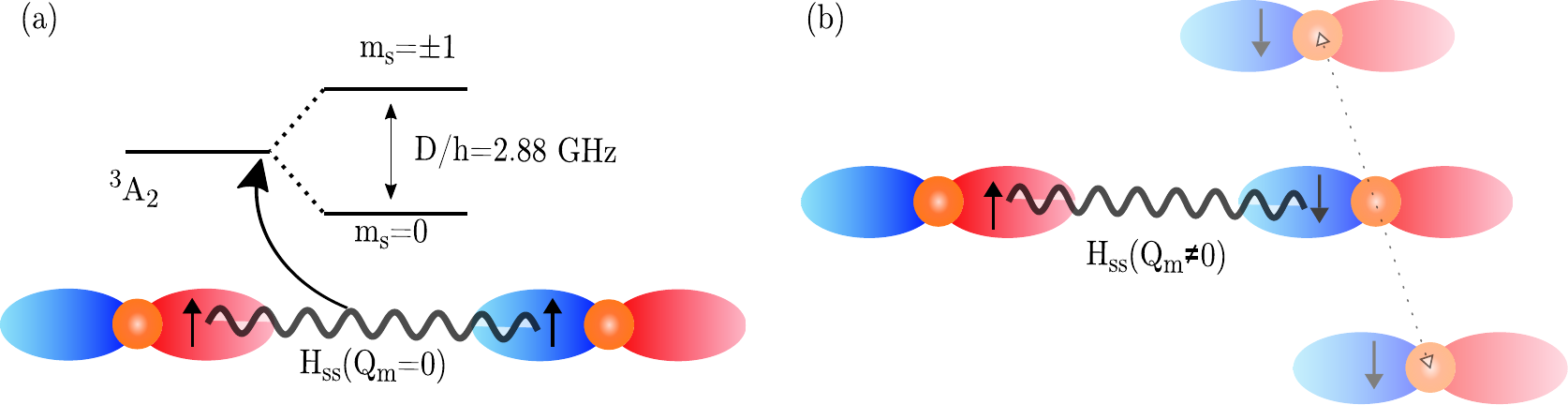}
	\caption{
	(a) In the static case (no phonons in the system) the spins ($\uparrow/\downarrow$) of the $\mathrm{NV}^-$ center interact via the static spin-spin interaction (wiggly line), which is responsible for the fine structure splitting of the $^3A_2$ ground state. (b) In the case the electron positions are coupled to phonons (right side of (b)), $\mathrm{H}_{\mathrm{ss}}$ becomes phonon dependent and can induce a spin flip. This allows to exchange energy between the spin system and the lattice and therefore allows the spin-system to equilibrate with the phonon bath.} 
	\label{fig:mechanism}
	 \end{center}
\end{figure}

\onecolumngrid
The idea to couple the spins to the phonons starting from the dipolar spin-spin interaction goes back to Waller \cite{Waller1932} and was the first impact on spin-lattice relaxation in general. \textcolor{black}{It was later neglected because most of the systems show a spin-orbit driven spin relaxation\cite{VanVleck1940,VanVleck1941,Overhauser1953,Elliott1954,Yafet1963}. The $\mathrm{NV}^-$ center containing only low $Z$ elements has a small spin-orbit coupling and the ground state triplet fine structure is given by the spin-spin interaction, which motivates an investigation of the spin-spin interaction as the driving source of spin lattice relaxation.}   The relaxation mechanism is depicted in Fig.~\ref{fig:mechanism}. If a phonon is excited, the dipolar spin-spin interaction 
\begin{align}
H_{\mathrm{ss}}=\underbrace{-\dfrac{\mu_0 g_e^{2} \mu_B^{2}}{4\pi}}_{\coloneqq \alpha} \dfrac{3(\boldsymbol{r}_{ij} \cdot \boldsymbol{S}_{i})(\boldsymbol{r}_{ij} \cdot \boldsymbol{S}_{j})-(\boldsymbol{S}_i \cdot \boldsymbol{S}_j) \boldsymbol{r}_{ij}^{2} }{\left|\boldsymbol{r}_{ij}\right|^{5}}
\label{eq:H_ss}
\end{align}
between the $i$-th and $j$-th electron is altered, because the electronic distance vector $\boldsymbol{r}_{ij}$ depends on the displacements of the ions $\{\boldsymbol{Q}_m\}$. Here $\mu_0$ denotes the vacuum permeability, $g_\mathrm{e}$ the g-factor \cite{Loubser1978} of the electron which is close to that of a free electron in the $\mathrm{NV}^-$ center with a value of \num{2.0028}, $\mu_B$ is the Bohr magneton and $\boldsymbol{S}_i$ and $\boldsymbol{S}_j$ are the spin vectors of the $i$-th and $j$-th electron. In his original work Waller neglected the orbital character of the electrons and treated them as point sources located at the positions of ions. This assumption will be dropped in the following derivation. The change of the position of the electron with the ionic motion has to be modelled to couple the electronic spin vectors in $\mathrm{H}_{\mathrm{ss}}$ to the ionic movements. This is achieved by defining a region $\Omega$ around each ion, in which the electronic orbital follows the movement of the ions rigidly, resulting in the electronic distance vector
\begin{align}
\label{eq:r_ij(Q)}
\boldsymbol{r}_{ij}(\{\boldsymbol{Q}_{m}\})=  \boldsymbol{r}_{ij}(\{\boldsymbol{Q}_{m}=0\})+\sum\limits_{m}\boldsymbol{Q}_{m}\underbrace{\left(\Theta(\boldsymbol{r}_{i} \in \Omega_m)-\Theta(\boldsymbol{r}_{j} \in \Omega_m)\right)}_{\coloneqq\Delta\Theta^{m}_{ij}} , 
\end{align}
where $\Theta(\boldsymbol{r} \in \Omega)$ is $1$, if the electron is inside $\Omega$ and $0$ otherwise.
For our calculations we use the Wigner-Seitz cell for $\Omega$, dividing space geometrically.
Since the ionic displacements in the low temperature regime are very small (mean square displacements are in the order of \SI{e-4}{\angstrom^{-2}}), a Taylor expansion to first order in $\boldsymbol{Q}_{m}$ is sufficient to calculate the transition rates between a $m_\mathrm{s} = \pm 1$ and a $m_\mathrm{s} = 0$ state. Thus, the spin-phonon interaction reads:
\begin{equation}\label{eq:V_sph}
\begin{split}
   V_{\mathrm{s-ph}}(\{\boldsymbol{Q}_{m}\})=\sum\limits_{m}\boldsymbol{Q}_{m}\dfrac{\partial H_{\mathrm{ss}}}{\partial \boldsymbol{R}_{m}}
  =\alpha\sum\limits_{m}&\Delta\Theta^{m}_{ij}\Big(\dfrac{
3\left((\boldsymbol{Q}_m \cdot \boldsymbol{S}_{i})(\boldsymbol{r}_{ij} \cdot \boldsymbol{S}_{j})+
(\boldsymbol{r}_{ij} \cdot \boldsymbol{S}_{i})(\boldsymbol{Q}_m \cdot \boldsymbol{S}_{j})\right)
}
{\left|\boldsymbol{r}_{ij}\right|^{5}}\\
&-\dfrac{15(\boldsymbol{r}_{ij} \cdot \boldsymbol{S}_{i})(\boldsymbol{r}_{ij} \cdot \boldsymbol{S}_{j})}{\left|\boldsymbol{r}_{ij}\right|^{7}}(\boldsymbol{r}_{ij}\boldsymbol{Q}_m) +\dfrac{3(\boldsymbol{S}_i \cdot \boldsymbol{S}_j)}{\left|\boldsymbol{r}_{ij}\right|^{5}}(\boldsymbol{r}_{ij}\boldsymbol{Q}_m)\Big)
\end{split}
\end{equation}

To extract the relevant matrix elements responsible for a transition between the $^3A_2$ levels, the spin operators $\boldsymbol{S}$ are expanded in raising and lowering operators. The Hamiltonian in \eqref{eq:V_sph} contains terms $(\boldsymbol{a}\cdot \boldsymbol{S}_i)(\boldsymbol{b}\cdot \boldsymbol{S}_j)$, where $\boldsymbol{a}$ and $\boldsymbol{b}$ are elements of $\{\boldsymbol{r}_{ij},\boldsymbol{Q}_m \}$, and a term $\boldsymbol{S}_i \boldsymbol{S}_j$, which can be rewritten as
\begin{align}
\label{eq:spinflip}
&\hspace{1cm}(\boldsymbol{a}\cdot \boldsymbol{S}_i)(\boldsymbol{b}\cdot \boldsymbol{S}_j)=
(a^{x} S_{i}^x + a^{y} S_{i}^{y} + a^{z} S_{i}^z)(b^{x} S_{j}^x + b^{y} S_{j}^{y} + b^{z} S_{j}^z) =\nonumber \\
&= \underbrace{\dfrac{1}{2}\bigg( S_i^{+}(a^x -ia^y) S_j^zb^z + S_i^-(a^x+ia^y) S_j^zb^z
+ S_i^za^z S_j^+(b^x-ib^y) + S_i^za^z S_j^-(b^x+ib^y)\bigg)}_{\text{single spin flip events}} + \nonumber\\
&\phantom{=}+\dfrac{1}{4}\bigg( S_i^+(a^x-ia^y)S_j^+(b^x-ib^y)  + S_i^-(a^x+ia^y)S_j^-(b^x+ib^y) + \nonumber \\
&\phantom{=}+ S_i^+(a^x-ia^y)S_j^-(b^x+ib^y) + S_i^-(a^x+ia^y)S_j^+(b^x-ib^y)   \bigg) + S_i^za^zS_j^zb^z 
\end{align}
and
\begin{align}
\boldsymbol{S_i}\boldsymbol{S_j}=\dfrac{1}{2}(S_{i}^{+}S_{j}^{-}+S_{i}^{-}S_{j}^{+})+S_{i}^{z}S_{j}^{z}.
\end{align}
The only matrix elements, which can cause a transition in the ground state triplet, are those, that contain only a single raising or lowering operator and are underbraced in Eq.~\eqref{eq:spinflip}, the remaining terms account for double spin-flip or no spin-flip events. Taking only the spin-flip matrix elements of $V_{\mathrm{s-ph}}$ in Eq.~\ref{eq:V_sph} into account we obtain the spin-flip potential
\begin{align}
V_{\mathrm{s-ph}}^{\mathrm{flip}}=\alpha\sum\limits_{m}& \Delta\Theta^{m}_{ij} \Big(
\dfrac{
3\left( \big( S_{i}^{\pm}S_{j}^{z}+S_{i}^{z}S_{j}^{\pm} \big) \big( (r_{ij}^x \mp i r_{ij}^y) Q_m^z + (Q_m^x \mp i Q_m^y) r_{ij}^z  \big)  \right)
}
{2\left|\boldsymbol{r}_{ij}\right|^{5}} - \nonumber\\
&-\dfrac{15 \big( (S_i^\pm S_j^z + S_i^z S_j^\pm)  (r_{ij}^x \mp i r_{ij}^y) r_{ij}^z  \big) \big( \boldsymbol{r}_{ij}\boldsymbol{Q}_m \big)}
{2\left|\boldsymbol{r}_{ij}\right|^{7}} 
\Big)
\end{align}
Likewise the ionic displacements $\boldsymbol{Q}_m$ are written in second quantized form \cite{Mahan2000}
\begin{align}
\label{eq:Q_n}
\boldsymbol{Q}_{m}=i\sum\limits_{\boldsymbol{q},\rho}\sqrt{\dfrac{\hbar}{2M_mN\omega_{\boldsymbol{q},\rho}}}(a^{\dagger}_{-\boldsymbol{q},\rho}e^{i\omega_{\boldsymbol{q},\rho}t}+a_{\boldsymbol{q},\rho}e^{-i\omega_{\boldsymbol{q},\rho}t}) \boldsymbol{\epsilon}_{\boldsymbol{q},\rho}^{m} e^{i\boldsymbol{q}\boldsymbol{R}_m^0}.
\end{align}
Here $M_m$ denotes the mass of the ion, $N$ the number unit cells, $a^{\dagger}$/$a$ the raising/lowering operator, $\boldsymbol{\epsilon}_{\boldsymbol{q},\rho}^m$ the polarization vector of the $m-$th ion in the mode and $\boldsymbol{R}_m^0$ is the equilibrium position of the ion. Substituting Eq.~\eqref{eq:Q_n} into Eq.~\eqref{eq:V_sph} and taking $V_{\mathrm{s-ph}}$ as a time dependent perturbation of the system leads to a Fermi's golden rule expression for a transition between the $m_\mathrm{s} = \pm 1$ and the $m_\mathrm{s} = 0$ states. The overall transition rate $\Gamma_{f \leftarrow i}$ from an initial to a final electronic state is obtained by a summation of the matrix elements of all final phonon states obeying energy conservation
\begin{equation}
\label{eq:fermi1}
\begin{split}
\Gamma_{f \leftarrow i}&=\dfrac{2\pi}{\hbar} \sum\limits_{f} \big| \braket{\tilde{N}_f,m_s^f|V_{\mathrm{s-ph}}^{\mathrm{flip}}|N_{i},m_s^i} \big|^2  \delta(E_f - E_i - h \nu)  \\
&=\dfrac{2\pi}{\hbar}\sum\limits_{\boldsymbol{k},\lambda} \big| \bra{m_s^f} \alpha\sum\limits_{m} i\sum\limits_{\boldsymbol{q},\rho}\sqrt{\dfrac{\hbar}{2M_mN\omega_{\boldsymbol{q},\rho}}} e^{i\boldsymbol{q}\boldsymbol{R}^m_0}\Delta\Theta^{m}_{ij} \cdot  \\
&\phantom{=}\Big(\dfrac{
3 \big( S_{i}^{\pm}S_{j}^{z}+S_{i}^{z}S_{j}^{\pm} \big) \big( r_{ij}^x \mp i r_{ij}^y \big) \bra{\tilde{N}_{f}}(a^{\dagger}_{-\boldsymbol{q},\rho}+a_{\boldsymbol{q},\rho}) \epsilon_{\boldsymbol{q},\rho}^{m,z} \ket{N_{i}} 
}
{2\left|\boldsymbol{r}_{ij}\right|^{5}}+\dfrac{
3 \bra{\tilde{N}_{f}}(a^{\dagger}_{-\boldsymbol{q},\rho}+a_{\boldsymbol{q},\rho}) (\epsilon_{\boldsymbol{q},\rho}^{m,x} \mp i\epsilon_{\boldsymbol{q},\rho}^{m,y}) \ket{N_{i}} r_{ij}^z  \big)
}
{2\left|\boldsymbol{r}_{ij}\right|^{5}} - \\
&\phantom{=}-\dfrac{15 \big( (S_i^\pm S_j^z + S_i^z S_j^\pm)  (r_{ij}^x \mp i r_{ij}^y) r_{ij}^z  \big) \big( \boldsymbol{r}_{ij}  \bra{\tilde{N}_{f}} (a^{\dagger}_{-\boldsymbol{q},\rho}+a_{\boldsymbol{q},\rho})\boldsymbol{\epsilon}_{\boldsymbol{q},\rho}^{m} \ket{N_{i}} \big)}
{2\left|\boldsymbol{r}_{ij}\right|^{7}} \Big) \ket{m_s^i} \big|^2 \delta(E_f - E_i - h \nu) .
\end{split}
\end{equation}
The raising and lowering operators acting on the initial phononic state $\ket{N_{i}}$ give $\sqrt{N_{\mathrm{ph}}+1}$ and $\sqrt{N_{\mathrm{ph}}}$ as eigenvalues of the particular state, where $N_{\mathrm{ph}}$ is the occupation number of the phonons. We assume $N_{\mathrm{ph}}$ to be the thermal occupations following the Bose-Einstein distribution. Putting everything together and considering the fact that only phonons with a single frequency at the spin-transition energy $D$ can take part in this process, the transition rates for emission and absorption of a phonon $\Gamma_{f \leftarrow i}$ read
\begin{align}
\label{eq:fermi2}
&\Gamma_{f \leftarrow i}=\dfrac{\alpha^2}{\hbar} \dfrac{ [N_{\mathrm{ph}}+1]^{emission}  [N_{\mathrm{ph}}]^{absorption} }{2N\omega} \cdot \nonumber \\
&\sum\limits_{\boldsymbol{k},\lambda} \big| \bra{m_s^f}\sum\limits_{m}\sqrt{\dfrac{1}{M_m}} \Delta\Theta^{m}_{ij}e^{i\boldsymbol{k}\boldsymbol{R}^m_0} \Big(
\dfrac{
3 \big( S_{i}^{\pm}S_{j}^{z}+S_{i}^{z}S_{j}^{\pm} \big) \big( r_{ij}^x \mp i r_{ij}^y \big) \epsilon_{\boldsymbol{k},\lambda}^{m,z}
}
{2\left|\boldsymbol{r}_{ij}\right|^{5}}+  \nonumber\\
&+\dfrac{
3 (\epsilon_{\boldsymbol{k},\lambda}^{m,x} \mp i \epsilon_{\boldsymbol{k},\lambda}^{m,y}) r_{ij}^z  \big)
}
{2\left|\boldsymbol{r}_{ij}\right|^{5}} - 
\dfrac{15 \big( (S_i^\pm S_j^z + S_i^z S_j^\pm)  (r_{ij}^x \mp i r_{ij}^y) r_{ij}^z  \big) \big( \boldsymbol{r}_{ij}   \boldsymbol{\epsilon}_{\boldsymbol{k},\lambda}^{m} \big)}
{2\left|\boldsymbol{r}_{ij}\right|^{7}} \Big) \ket{m_s^i} \big|^2 \delta(\nu=\text{\SI{2.88}{\giga\hertz}})
\end{align}
To emphasize the temperature dependence of the relaxation rate this is rewritten as
\begin{align}
\label{eq:Gamma_fi}
\Gamma_{f \leftarrow i}= \begin{cases} (N_{\mathrm{ph}}+1) \Gamma_0 & \text{for emission of a phonon} \\ N_{\mathrm{ph}} \Gamma_0 & \text{for absorption of a phonon}\end{cases}
\end{align}
with $\Gamma_{0}$ being the transition rate at zero temperature. To simulate an ensemble of spins relaxing from a non-equilibrium spin-distribution to equilibrium with the environment, both deexcitations and excitations of spins have to be considered \cite{Scott1962} and the following rate equations for the occupations $N_{m_\mathrm{s}}$ have to be solved for our system with a degenerate excited state:
\begin{align}
\label{eq:rate_equations1}
\dot{N}_{m_\mathrm{s}\pm 1}&=-\Gamma_{0}(N_{\mathrm{ph}}+1) N_{m_\mathrm{s}\pm 1} + 2\Gamma_0 N_{\mathrm{ph}} N_{m_\mathrm{s}=0} \nonumber \\
\dot{N}_{m_\mathrm{s}=0}&=-\dot{N}_{m_\mathrm{s}\pm 1} \nonumber \\
\end{align}
The solution is straightforward by introducing the occupation difference $\Delta N = N_{m_\mathrm{s} = \pm 1} - N_{m_\mathrm{s} = 0}$, since it obeys a simple exponential decay law to its thermal equilibrium value $\Delta N_{\mathrm{th}}$ according to 
\begin{align}
\label{eq:rate_equations2}
\dfrac{d}{dt}\Delta N=-\underbrace{(3N_{\mathrm{ph}}+1)\Gamma_0}_{\Gamma_1=1/T_1}(\Delta N-\Delta N_{\mathrm{th}}).
\end{align}
The calculation of the zero temperature transition rate $\Gamma_0$ between the $^3A_2$ sublevels is sufficient to extract the transition rates in the low temperature regime, where single phonon processes are dominating over two phonon-processes.
\twocolumngrid
\section{Methods}
\label{sec:Methods}
\begin{figure}
	 \begin{center}
	 \includegraphics{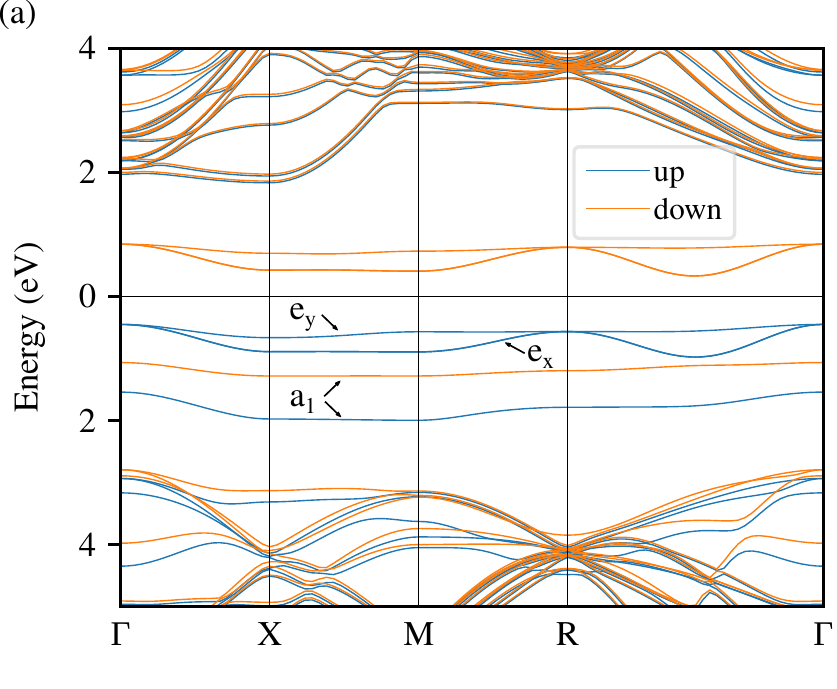}
	 \includegraphics{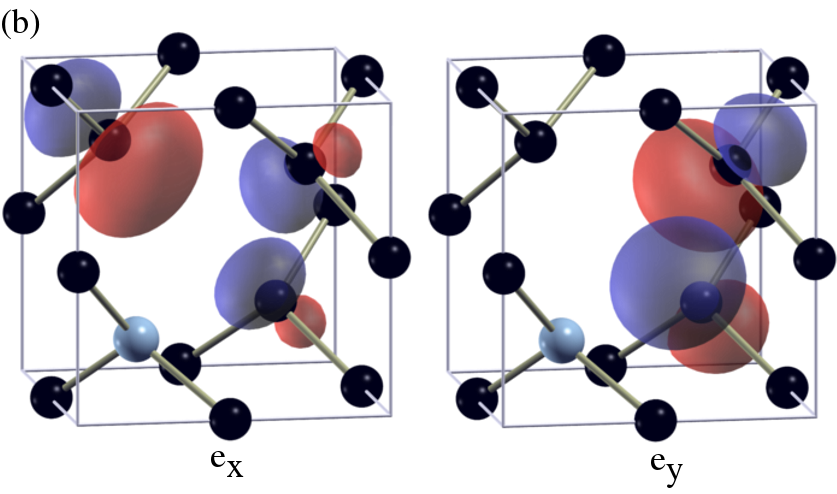}
	\caption{
	(a)The spin-polarized bandstructure calculated with HSE in a supercell containing 64 atoms. 4 electrons of the $\mathrm{NV}^-$ center are located inside the bandgap and occupy the three orbitals $a_1$,$e_\mathrm{x}$ and $e_\mathrm{y}$. The blue lines denote spin up bands, the orange ones spin down bands. The electrons in the $e_\mathrm{x}$ and $e_\mathrm{y}$ orbitals are responsible for spin-polarization. (b) The isosurfaces of the spin-polarized $e_\mathrm{x}$ and $e_\mathrm{y}$ orbitals. The maximally localized orbitals $c_1$, $c_2$, $c_3$ and $n$ on the 4 adjacent atoms next to the vacancy were added up to fulfill the $C_{3v}$ symmetry constraints \cite{Loubser1978,Lenef1996} resulting in $e_\mathrm{x}  \propto  2c_3-c_1-c_2$ and $e_\mathrm{y} \propto c_1-c_2$. It is evident, that the spin density is mainly located at the carbon atoms adjacent to the vacancy.
	} 
	\label{fig:orbitals}
	 \end{center}
\end{figure}

The calculation of $\Gamma_0$ requires the spin-polarized electronic orbitals as well as the phononic bandstructure, density of states and the polarization vectors for all the modes. We perform \textit{ab initio} calculations using density functional theory  on supercells containing 64,128 and 512 lattice sites with one $\mathrm{NV}^-$ center employing the Vienna Ab initio Simulation Package (VASP \cite{Kresse1996}) using projector augmented wave pseudopotentials \cite{Kresse1999}. We use the local density approximation and a generalized gradient approximation included in the PBE \cite{Perdew1996} exchange correlation potential for structural relaxations and force calculations. The electron properties in the relaxed structure are also calculated using the SCAN \cite{Sun2015} and HSE functional \cite{Heyd2003,Krukau2006}. Plane waves up to a cutoff of \SI{700}{\electronvolt} are included and the first Brillouin zone is sampled with a \SI{4x4x4}{} Monkhorst Pack grid\cite{Monkhorst1976}. A subtle relaxation of the ions resulting in forces on the atoms of less than \SI[per-mode=fraction]{1}{\milli\electronvolt \per \angstrom} shows that the neighbouring carbon atoms and the nitrogen atom move away from the vacancy, where the nitrogen atom is further displaced in accordance with an earlier study \cite{Gali2008}. Since we are interested in the spin-polarized orbitals, we use the relaxed positions to calculate the electronic band structure. It is found that the $a_1$,$e_\mathrm{x}$ and $e_\mathrm{y}$ orbitals are located inside the bandgap and that $e_\mathrm{x}$ and $e_\mathrm{y}$ are the spin-polarized orbitals (see Fig.~\ref{fig:orbitals}a). This familiar result \cite{Loubser1978,Goss1996,Lenef1996,Gali2008} allows to extract these orbitals by applying the wannier90 package \cite{Marzari2012,Mostofi2014} to obtain the maximally localized orbitals on the nearest neighbour atoms of the vacancy. By considering the symmetry of the defect we add up the maximally localized orbitals to fulfill the $C_{\mathrm{3v}}$ symmetry constraints and end up with the spin polarized $e_\mathrm{x}$ and $e_\mathrm{y}$ orbitals of the $^3A_2$ groundstate shown in Fig.~\ref{fig:orbitals}b.
Building Slater determinants with these orbitals we calculate the electronic matrix elements $\braket{m_\mathrm{s}^f|\dfrac{r_{ij}^x \mp i r_{ij}^y}{\left| \boldsymbol{r}_{ij}\right|^5}|m_\mathrm{s}^i}$, $\braket{m_\mathrm{s}^f|\dfrac{r_{ij}^z}{\left| \boldsymbol{r}_{ij}\right|^5}|m_\mathrm{s}^i}$, $\braket{m_\mathrm{s}^f|\dfrac{(r_{ij}^x \mp i r_{ij}^y)r_{ij}^z r_{ij}^{x/y/z}}{\left| \boldsymbol{r}_{ij}\right|^7}|m_\mathrm{s}^i}$, which occur for every phononic polarization in Eq.~\eqref{eq:Gamma_fi}.\\
The phonons are modelled by using the small displacement method within the harmonic approximation similar to a previous study \cite{Gali2011}. We use the PHONOPY package \cite{Togo2015} to extract the necessary displacements to build the dynamical matrix and apply it to the diagonalization thereof. We sample the Brillouin zone with a very dense mesh to extract \num{10000} phonon polarization vectors at the transition frequency per band and the respective local group velocities for any particular $\boldsymbol{k}$-point. The phononic density of states is calculated according to a Debye-model where we take the $\boldsymbol{k}$-dependence of the group velocity into account. With the polarization vectors $\{\boldsymbol{\epsilon}_{\boldsymbol{k},\lambda}^m\}$ and the density of states in hand, we have sufficient data to perform the summation over all the final phonon modes and calculate the numerical value of the spin-lattice relaxation rate $\Gamma_0$. 
\section{Results}
\label{sec:Results}
\begin{figure}
	 \begin{center}
	 \includegraphics{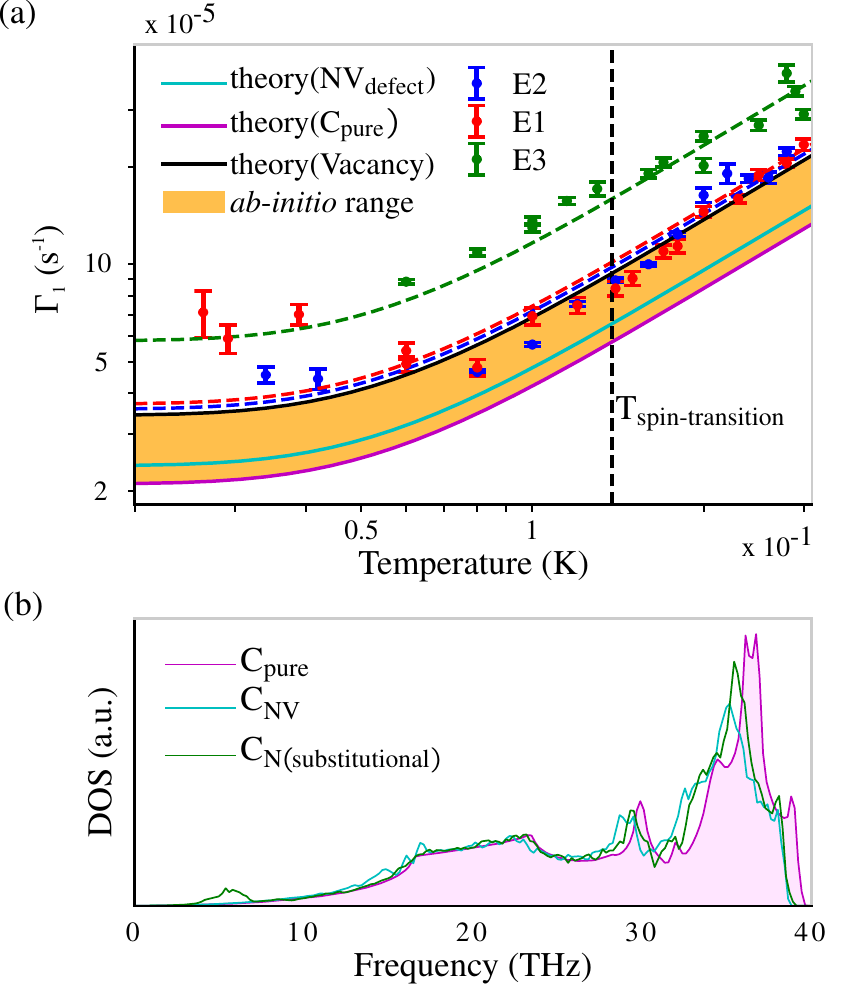}
	\caption{(a) The measured spin-lattice relaxation rates for 3 different samples ($E1$, $E2$ and $E3$ [data taken from \cite{Astner2017}]). Dashed lines are least square fits for the temperature dependence according to Eq.~\eqref{eq:rate_equations2}. \textcolor{black}{The theoretical results (yellow range) depend on the phononic density of states: The magenta line denotes the calculated relaxation rate, if a Debye-model with the velocity of sound of a pure diamond is applied, the cyan line represents the case, where the $\boldsymbol{k}$-dependent sound velocity of the simulated cell with 1 $\mathrm{NV}^-$ center was used and the black line results for the DOS of a 64 lattice site diamond cell containing a vacancy.} (b) Difference in the phononic density of states between a perfect diamond crystal (red) and diamond crystals with point defects. If a point defect is present the DOS is shifted towards lower energy excitations.
	} 
	\label{fig:rates}
	 \end{center}
\end{figure}
After carrying out the calculations, we end up with a theoretically predicted temperature dependent relaxation rate $\Gamma$, which can be compared to the experiment. As shown in Fig.~\ref{fig:rates}a we find a direct single phonon process \cite{Scott1962,Shrivastava1983} at temperatures above the spin-transition $T \gg D / k_B$ where thermal phonons excite and deexcite the spins by induced emission or absorption resulting in a linear dependence of $\Gamma$ on $T$. This temperature dependence stems from the high temperature limit of the Bose-Einstein distribution, where $N_{\mathrm{ph}} \propto T$ in Eq.~\eqref{eq:rate_equations2}. At temperatures below the spin-transition the \SI{2.88}{\giga\hertz} phonons start to freeze out and the only decay channel left for a spin-transition is the temperature independent spontaneous emission of a phonon occuring with a rate $\Gamma_0$ which results in the observed plateau in the low temperature regime.To compare the calculated rates with experiment, the treatment of the samples has to be explained: To create $\mathrm{NV}^-$ centers in diamond, samples with a high initial nitrogen concentration (type Ib diamond) are irradiated by electrons, neutrons or ions in order to obtain vacancies followed by an annealing procedure \cite{Davies1976,Mainwood1994,Nobauer2013}. The influence of the radiation damage on the phononic density of states is essentially unknown, but irradiation will create point defects, which can shift the phononic density of states towards lower energy excitations \cite{Bedoya-Martinez2016}. We simulate this effect and calculate the density of states for diamond crystals with defects and compare them with a perfect crystal. Introducing point defects (substitutional nitrogens and vacancies) in the diamond structure the phononic density of states indeed shifts towards lower frequencies as illustrated in Fig.~\ref{fig:rates}b . However, we can only estimate the phononic density of states in the irradiated crystals. We model the phonons using the phononic densities of states for the simulated cells and the calculated relaxation rates $\Gamma_{0,\textit{ab initio}}=\text{\SIrange{2E-5}{3E-5}{\per \second}}$ are close to the lowest experimental values ($\Gamma_{0,exp}=\SI{3.47+-0.16e-5}{\per \second}$) \cite{Astner2017}. 
\section{Conclusion}
\label{sec:Conclusion}
In this paper we have shown that the very low spin-lattice relaxation rates of the $\mathrm{NV}^-$ center in diamond can be explained by the change of the dipolar spin-spin interaction induced by the movement of the ions as proposed originally by Waller in 1932 \cite{Waller1932}. We coupled the electronic distance vector $\boldsymbol{r}_{ij}$ to the ionic movement by a first order Taylor expansion in the ionic displacement vectors $\{\boldsymbol{Q}_m\}$ and ended up with an effective spin-phonon interaction $V_{\mathrm{s-ph}}$. We applied this interaction as a perturbation of the system to calculate the transition rates between the $^3A_2$ ground state spin triplett \textit{ab initio} using density functional theory by modelling the electronic wavefunctions and the phonons in a supercell containing 512 atoms. The calculated relaxation rates are comparable to the ones measured for samples that show little crystal damage. We propose that the deviation to samples with a strong irradiation damage is caused by the difference in the phononic density of states due to the irradiation treatment.
Knowing the fundamental mechanism of spin-phonon interaction in this system will allow us to further investigate on higher order two phonon Raman processes and Orbach processes at higher temperatures for the spins of the $\mathrm{NV}^-$ center in diamond. In this work we show that the predictive power of modern \textit{ab initio} calculations allows the identification of critical phonon modes, which could lead to tailoring the relaxation time in future applications.                                                                                                                                                                                                                                                     


\begin{acknowledgments}
J.G. and P.M. were supported by the FWF SFB VICOM (Project F4109-N28), T.A. and A.A. acknowledge support by the Doctoral School Building Solids for Function (FWF Project W1243), J.M. has been supported by the TOP grant of TU Wien and J.S. by the WWTF project SEQUEX (Project MA16-066).
\end{acknowledgments}
\bibliography{Main.bib}

\end{document}